\documentclass[aps,prx,twocolumn,showpacs,longbibliography]{revtex4-1}
\usepackage{amsmath}
\usepackage{graphicx}
\usepackage[colorlinks,linkcolor=blue,citecolor=blue,urlcolor=green]{hyperref}
\usepackage{stix}
\usepackage[dvipsnames]{xcolor}

\newcommand{\iu}{\mathrm{i}}
\newcommand{\dd}{\mathrm{d}}

\renewcommand{\vec}[1]{\mathbf{#1}}
\newcommand{\gvec}[1]{\boldsymbol{#1}}
\newcommand{\en}{\varepsilon}
\newcommand{\hc}{\hat{c}}
\newcommand{\hcd}{\hat{c}^\dagger}
\newcommand{\hd}{\hat{d}}
\newcommand{\hdd}{\hat{d}^\dagger}

\begin{document}

\title{Nonthermal 
switching of charge order: 
dynamical slowing down 
and optimal control}

\author{Michael Sch\"uler}
\author{Yuta Murakami}
\author{Philipp Werner}
\affiliation{Department of Physics, University of Fribourg, 1700 Fribourg, Switzerland}

\begin{abstract}
  We investigate the laser-induced dynamics of electronically driven charge-density-wave
  order. A  comprehensive mean-field analysis of the attractive Hubbard model in the 
  weak-coupling regime reveals ultrafast switching and ultrafast melting of the order
  via a nonthermal pathway. The resulting nonequilibrium
  phase diagram exhibits multiple dynamical phase transitions
  with increasing field strength.  Using an intuitive pseudospin picture, we
  show that the laser can be regarded as a external (pseudo) magnetic
  field, and that the distinct dynamical regimes can be connected to the 
  spin precession angle.  We furthermore study the effects of
  electron-electron interactions beyond mean-field to show that the
  main features of the phase diagram are robust against scattering or thermalization
  processes.  For example, the nonthermal state with switched order
  is characterized by a particularly slow relaxation.  We also
  demonstrate how these nonthermal phases can be stabilized by
  tailoring the pulse shape with optimal control theory.
\end{abstract}

\pacs{}

\maketitle

The field of ultrafast coherent control of symmetry-broken phases in
condensed matter experiences a surge of research activity.  Important
examples include the transient enhancement of
superconductivity~\cite{mankowsky_nonlinear_2014,kaiser_optically_2014,
  mitrano_possible_2016,sentef_theory_2016,kennes_transient_2017,babadi_theory_2017,
murakami_nonequilibrium_2017,mazza_nonequilibrium_2017,bittner_light-induced_2017},
light-controlled order of excitonic
condensates~\cite{mohr-vorobeva_nonthermal_2011,golez_photoinduced_2016,
  mor_ultrafast_2017,murakami_photo-induced_2017} or the transient
manipulation of
spin~\cite{okamoto_ultrafast_2006,conte_snapshots_2015},
orbital~\cite{cavalleri_coherent_2007} and charge
order~\cite{onda_photoinduced_2008,huber_coherent_2014,shen_nonequilibrium_2014,
  caviezel_time-dependent_2014}.  Since charge-density-wave (CDW) order directly
couples to external electric fields, it provides an ideal
platform for inducing and tracing photoinduced phase transitions.  Several recent
experiments have addressed the mechanism behind the CDW-to-metal transition
by ultrashort laser pulses. If the CDW originates from a
lattice distortion, the light-induced melting is
accompanied by coherent phonon
excitations~\cite{cavalleri_evidence_2004,chollet_gigantic_2005,okamoto_ultrafast_2006,
  cavalleri_coherent_2007,
  onda_photoinduced_2008,huber_coherent_2014}. In contrast, a CDW
induced by electron-electron (e--e) correlations 
can be melted on the femtosecond
timescale~\cite{iwai_photoinduced_2007,yusupov_single-particle_2008,
  mohr-vorobeva_nonthermal_2011,porer_non-thermal_2014}, and
nonthermal melting has been
observed~\cite{mohr-vorobeva_nonthermal_2011,porer_non-thermal_2014}.
These developments have stimulated intensive theoretical research of the
laser-driven CDW
dynamics~\cite{yonemitsu_photoinduced_2007,shen_nonequilibrium_2014,
hashimoto_photo-induced_2014,matveev_time-domain_2016,hwang_optimizing_2016,
wang_using_2016,hashimoto_photoinduced_2017}.
Furthermore, since CDW is typically
competing with superconductivity~\cite{kamihara_iron-based_2006}, the suppression of
such orders provides a route to light-induced
superconductivity~\cite{fausti_light-induced_2011}.

From a fundamental perspective, universal features in the time
evolution of ordered states are attracting great interest. The dynamics after a generic excitation can exhibit 
qualitative changes -- such as transitions from transiently trapped to vanishing 
order -- beyond some critical excitation strength, which cannot be 
explained in terms of the total energy of the system within an 
equilibrium picture. Such dynamical
phase transitions~\cite{eckstein_thermalization_2009, heyl_dynamical_2014,zunkovic_dynamical_2016} 
have been 
studied for various kinds of ordered phases such as
superconductors~\cite{yuzbashyan_dynamical_2006,barankov_synchronization_2006,peronaci_transient_2015},
excitonic insulators~\cite{murakami_photo-induced_2017},
antiferromagnetic~\cite{schiro_time-dependent_2010,werner_nonthermal_2012,tsuji_nonthermal_2013},
ferromagnetic~\cite{zunkovic_dynamical_2016,lerose_chaotic_2017} and bosonic~\cite{sciolla_quantum_2010} systems.

Here we show that the e--e driven CDWs excited with short
pulses can also exhibit such dynamical phase transitions. In contrast to
the usual quench scenario, we identify \emph{multiple} regimes of
ultrafast nonthermal melting or switching depending on the pulse
strength. We present a corresponding \emph{nonequilibrium phase
  diagram} which reveals nonthermal slowing down and long-lived
nonthermal states despite increased energy absorption.  We provide an
intuitive understanding of the switching and melting behavior based on
time-dependent mean field (MF) theory and a pseudospin picture.  In
addition, by taking e--e scattering into account at the local
second-Born level, we demonstrate that the qualitative features
identified in the MF dynamics are also present in a more sophisticated
description, which captures thermalization effects.  To access these
interesting transient states efficiently, we employ quantum optimal
control theory
(QOCT)~\cite{eitan_optimal_2011,reich_monotonically_2012,schmidt_optimal_2011},
which provides optimized laser pulses for either switching or
suppressing the CDW order. By minimizing energy absorption
, we can suppress heating effects and stabilize the nonthermal
states.

\begin{figure}[b]
  \includegraphics[width=\columnwidth]{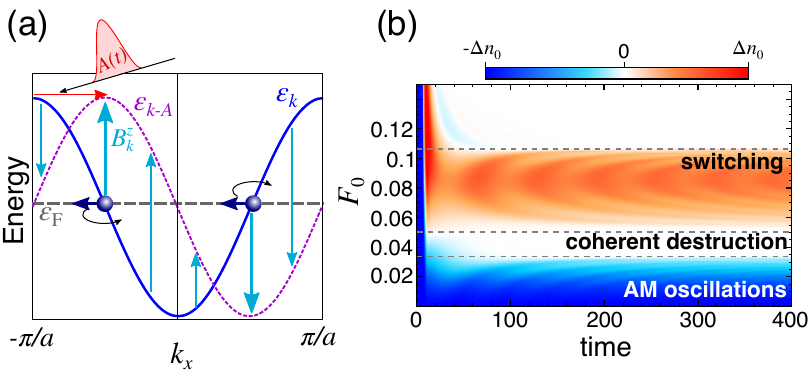}
  \caption{(a) Schematic of the switching dynamics in the pseudospin
    picture (see text). A single-cycle field pulse shifts the band
    structure, inducing a $k$-dependent effective magnetic field
    $B^z_k$. By tuning the precession time of the pseudospins, the
     contribution arising from the Fermi level can be switched
    completely. (b) MF dynamics of the order parameter induced by a
    single-cycle pulse as a function of its amplitude $F_0$. 
\label{fig:intro}}
\end{figure}

{\it Model--} In this work, we study the 
attractive Hubbard model at half filling
\begin{align}
  \label{eq:Ham1}
  \hat{H} &= -J_0\sum_{\langle ij\rangle, \sigma} \hat c^\dagger_{i\sigma} \hat c_{j\sigma} +
  \frac{U}{2}\sum_{i} \left(\hat{n}_{i\sigma}-\frac12\right)
  \left(\hat{n}_{i\bar\sigma}-\frac12\right) \ ,
\end{align}
where $\langle i j\rangle$ denotes nearest neighbors, $U$ is the
on-site interaction and $\hat{n}_{i \sigma} = \hat c^\dagger_{i\sigma}
\hat c_{i\sigma}$.  This model gives rise to CDW order and
superconductivity. Here we focus on the CDW state, which is
characterized by the broken symmetry between neighboring sublattice
sites. 
In what follows we focus on the weak coupling regime and treat a
one-dimensional (1D) configuration to simplify the numerical calculations,
having in mind a quasi 1D system embedded in higher dimensions~\cite{Note1}.
%

In momentum space, the kinetic term of the Hamiltonian~\eqref{eq:Ham1}
reads 
$\hat{H}_\mathrm{kin} = \sum_{k,\sigma} \en_{k} \hcd_{k \sigma}\hc_{k
  \sigma}$ with the free band structure $\en_k = -2J_0 \cos(k a)$
(lattice constant $a$). The time-dependent external laser fields,
described by the vector potential $A(t)$, are incorporated by the
Peierls substitution $\en_k(t) = \en_{k-A(t)}$.  To allow for CDW
order, one introduces the nesting vector $Q=\pi/a$, which leads to
nested bands and a reduced Brillouin zone.  In what follows, we
measure energies in units of $J_0$ and time in units of
$J^{-1}_0$, while $k$ is given in units of $a^{-1}$. 
We use the e--e interaction $U=-2$, 
which corresponds to the weak-coupling regime.

{\it Results--}
First we discuss the MF dynamics which provides valuable insights into the
short-time dynamics of the CDW system. 
To this end, we introduce the pseudospin representation by
$\hat{S}^\alpha_{k\sigma} =
\hat{\vec{c}}^\dagger_{k\sigma}\hat{\sigma}^\alpha
\hat{\vec{c}}_{k\sigma} /2 $ (Pauli matrices
$\hat{\sigma}^\alpha$) with $\hat{\vec{c}}_{k\sigma} =
(\hc_{k\sigma},\hc_{k+Q,\sigma})$. 
This concept has provided an intuitive understanding of related
problems~\cite{anderson_random-phase_1958,tsuji_nonthermal_2013,
matsunaga_light-induced_2014,sentef_theory_2015,murakami_photo-induced_2017}.
The CDW order parameter describing the charge difference with respect
to $(k,k+Q)$ (or, equivalently, sublattice sites) is given by
$\Delta n = (1/N_k) \sum_{k\sigma} \langle \hcd_{k+Q,\sigma}
\hc_{k\sigma} \rangle = (1/N_k)\sum_{k,\sigma}\langle
\hat{S}^{x}_{k\sigma} \rangle$.  The current between the sublattice
sites is, on the other hand, related to
$\langle \hat{S}^y_{k\sigma} \rangle$.  CDW order corresponds
to a finite total pseudospin projection in the $x$-direction, while
the normal state in equilibrium consists of spins along the $z$-direction.

With these definitions, the MF Hamiltonian and the corresponding
equation of motion is cast into the simple form
\begin{align}
  \label{eq:Ham_pseudo}
  \hat{H}^\mathrm{MF}(t) = \sum_{k\sigma}
  \vec{B}_k(t)\cdot\hat{\vec{S}}_{k\sigma} \ ,  \
  \frac{\dd}{\dd t} \hat{\vec{S}}_{k\sigma} = \vec{B}_k(t) \times
  \hat{\vec{S}}_{k\sigma} \ ,
\end{align}
where the momentum dependent pseudo-magnetic field is given by
$B^x_{k}(t) = U \Delta n(t)$, $B^y_{k} = 0$ and
$B^z_k(t) = \en_k(t)-\en_{k+Q}(t)$. 

The pseudo-spin representation~\eqref{eq:Ham_pseudo} provides an intuitive picture for the
control of CDW configurations by external fields.  
For instance, the
asymmetric electronic charge on the sublattice sites can be switched
with a short single-cycle pulse (SCP) with a pulse duration
$T_\mathrm{p}$, see Fig.~\ref{fig:intro}(a).  During the pulse, the
kinetic energy of the electrons with momentum $k$ is shifted by the
vector potential to $\epsilon_{k- A(t)}$.  In the pseudo-spin picture,
this acts as a momentum-dependent magnetic field in the $z$ direction,
whose maximum strength is half the band width.  
In the weak coupling regime, the major contribution to the
CDW comes from the Fermi surface. Furthermore, $B^x_{k}(t)$ is
smaller than the band width and can be neglected during the pulse.
Therefore, the CDW dynamics can be regarded as spin
precession around the $z$ direction.

By tuning the pulse amplitude and $T_\mathrm{p}$ we can control the
pseudospins at the two Fermi points , which rotate in opposite
directions (Fig.~\ref{fig:intro}(a)).  Optimizing the pulse to induce
a precession by $(2n+1)\pi$ ($n \in \mathbb{N}$), the sign
of 
the order is switched.  Inducing a rotation by $\frac{(2n+1)}{2}\pi$
($n \in \mathbb{N}$), on the other hand, the pseudospin projections
cancel out, which leads to efficient destruction of the order.  Note
that the pseudospins at different $k$ precess under a $k$-dependent
pseudo-magnetic field. Hence the complete pseudospin dynamics is
subject to dephasing, which can reduce the size of the order parameter
after the switching.

We now proceed to the numerical investigation of the
pulse-induced CDW dynamics.  We propagated the MF
Hamiltonian~\eqref{eq:Ham_pseudo} in the presence of a SCP
described by the electric field
\mbox{$E(t) = F_0 \sin^2[\pi(t-t_0)/T_\mathrm{p}]\sin[\omega(t-t_0)]
  $} with $T_\mathrm{p}=2\pi/\omega$ (vector potential
$A(t)=-\int_0^t dt'E(t')$).  We found that $T_\mathrm{p}=13.6$ matches
the typical time scale of the system and allows efficient switching as in
the pseudospin picture.

In our MF study, we use the renormalized parameters
$\widetilde{J}_0 = 0.89 J_0$ and $\widetilde{U}=0.625 U$. For this
choice, the MF band structure provides a good approximation of the
renormalized bandstructure~\cite{supplement}, and thus allows
us to use the MF dynamics as a reference for the correlated treatment below. In both
cases, one obtains an equilibrium order parameter of
$\Delta n_\mathrm{eq} \equiv -\Delta n_0 = -0.121$. The inverse
temperature is fixed at $\beta = 40$. 

\begin{figure}[t]
  \includegraphics[width=\columnwidth]{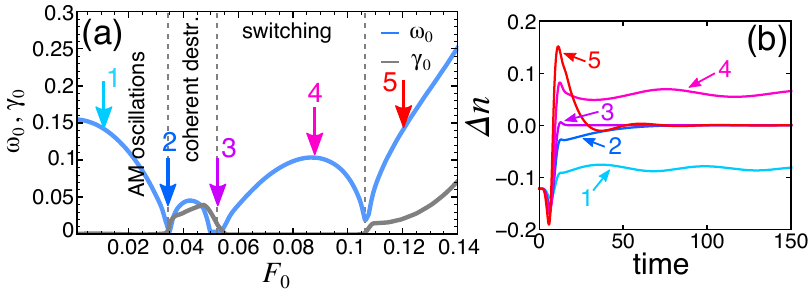}
  \caption{(a) Nonequilibrium phase diagram depicting the
    oscillation frequency $\omega_0$ and damping constant $\gamma_0$
    of the time-dependent order parameter. (b) Dynamics of $\Delta
    n(t)$ at the values of $F_0$ indicated in (a).
\label{fig:mfneq}}
\end{figure}

The MF dynamics of $\Delta n$ for varying pulse strengths $F_0$ is
presented in Fig.~\ref{fig:intro}(b).  Several regimes of
qualitatively distinct dynamics can be identified (multiple dynamical
phase transitions).  For weak fields, the system exhibits persistent
amplitude mode (AM) oscillations around a mean value $\Delta
\bar{n}$. The corresponding frequency decreases with decreasing
$|\Delta \bar{n}|$. The second regime corresponds to a rapid
destruction of the CDW order. In particular, for $F_0 \simeq 0.05$ the
order is almost immediately destroyed after the pulse, which is due to
dephasing of the pseudospins and does not correspond to a thermal
melting process.  We therefore refer to this regime as coherent
destruction (CD).
Increasing $F_0$ further, the order is switched,
as expected from the pseudospin picture. One observes the
emergence of AM oscillations around the switched
value. However, a complete switching is prevented by the dephasing of the
pseudospins. The order parameter reaches an average value of 
$\Delta\bar{n}\simeq 0.66 \Delta n_0$.

Let us also comment on the situation where an electron-phonon coupling
(e--ph) contributes to the CDW. The analogous MF analysis of the
Hubbard-Holstein model reveals~\cite{supplement} a similar
nonequilibrium phase diagram in the case where the e--e interactions
dominate the CDW formation. For dominant e--ph coupling, in contrast,
the dynamics is characterized by persistent coherent phonon
excitations which inhibit the long-time stable switching or
destruction of the CDW~\footnote{Note2}.  Hence, the distinct regimes
in Fig.~\ref{fig:intro}(b) are a clear indication of
correlation-driven charge order.

For a quantitative discussion of the different regimes we analyze the
dynamics by fitting the order parameter to the damped
oscillating function
$f(t) = c + [a \cos(\omega_0 t)+b\sin(\omega_0t)]e^{-\gamma_0 t}$ in
the long-time limit. This form describes (i) damped AM oscillations and
(ii) non-oscillatory ($\omega_0 =0$) decay of the order.  The
extracted damping constants $\gamma_0$ and oscillation frequencies
$\omega_0$ are shown as a function of field amplitude in Fig.~\ref{fig:mfneq}(a).  The regime of AM
oscillations is characterized by a finite $\omega_0$ and a very small
$\gamma_0$. A closer inspection reveals a very slow algebraic decay.
The CD of the order is found in the region $0.035 < F_0 < 0.05$. It is
bounded by two special points of slowing down at nonthermal critical
points \cite{tsuji_nonthermal_2013}, where oscillations of the order parameter are fully suppressed
(Fig.~\ref{fig:mfneq}(b)). In particular, at the right boundary
$F_0 = 0.05$ the order is 
strongly suppressed already a short time after the pulse ("sweet
spot", marked as line 3 in Fig.~\ref{fig:mfneq}(b)). Within the CD
regime, the system exhibits strongly damped AM oscillations after a
rapid reduction of the mean value of $\Delta n$.
Increasing the field strength further, the CDW system enters the
switching regime. It is again characterized by AM oscillations, but
around the switched value of the order parameter.  For even larger
pulse amplitude, the order is again destroyed and a second CD regime
emerges.

Now we consider the effect of scattering and thermalization processes
on the nonequilibrium phase diagram.  In general, 
the absorbed pulse energy ($E_\mathrm{abs}$) will lead to a thermal melting of the order in
the long-time limit in most scenarios. These heating effects, which
are often ignored in theoretical studies, deserve proper attention in
the discussion of light-controlled
order~\cite{murakami_nonequilibrium_2017}. 
To address this, we switch to a description in terms of nonequilibrium Green's functions
~\cite{balzer_nonequilibrium_2012,stefanucci_nonequilibrium_2013}. The
single-particle Green's function (GF) is defined as
$\vec{G}_{\sigma}(k;z,z^\prime) = -\iu \langle \mathcal{T}
\hat{\vec{c}}_{k\sigma}(z) \hat{\vec{c}}^\dagger_{k\sigma}(z^\prime)
\rangle$, where the arguments $z$, $z^\prime$ refer to the L-shaped
Matsubara-Keldysh contour $\mathcal{C}$ and $\mathcal{T}$ denotes the
contour ordering operator.
Given some approximation to the
self-energy $\gvec{\Sigma}_\sigma(k;z,z^{\prime})$,
the Kadanoff-Baym equations (KBEs) yield the interacting
time-dependent GF, from which the relevant observables can be
extracted~\cite{supplement}. 

Generally, the self-energy is expressed as a diagrammatic
series. Since the e--e interactions are weak in our setup, the
second-order (second-Born, 2B) approximation to the self-energy 
yields an accurate description. In full generality, it incorporates
momentum-dependent scattering processes.  For low dimensional systems,
however, the available phase space is strongly restricted and the e--e
scattering is strongly suppressed.  For numerical simplicity and to
partially reflect that the system is embedded in higher-dimensions we
employ the local 2B (2Bloc) approximation
$\Sigma_{ij,\sigma}(z,z^\prime) = \delta_{ij} U^2
G_{ii,\sigma}(z,z^\prime) G_{ii,\bar{\sigma}}(z,z^\prime)
G_{ii,\bar{\sigma}}(z^\prime,z)$.  The e--e scattering may thus be
overestimated compared to an actual quasi-1D system. Hence, one can
expect that the 2Bloc approximation provides an upper bound for
thermalization effects.

\begin{figure}[t]
  \includegraphics[width=\columnwidth]{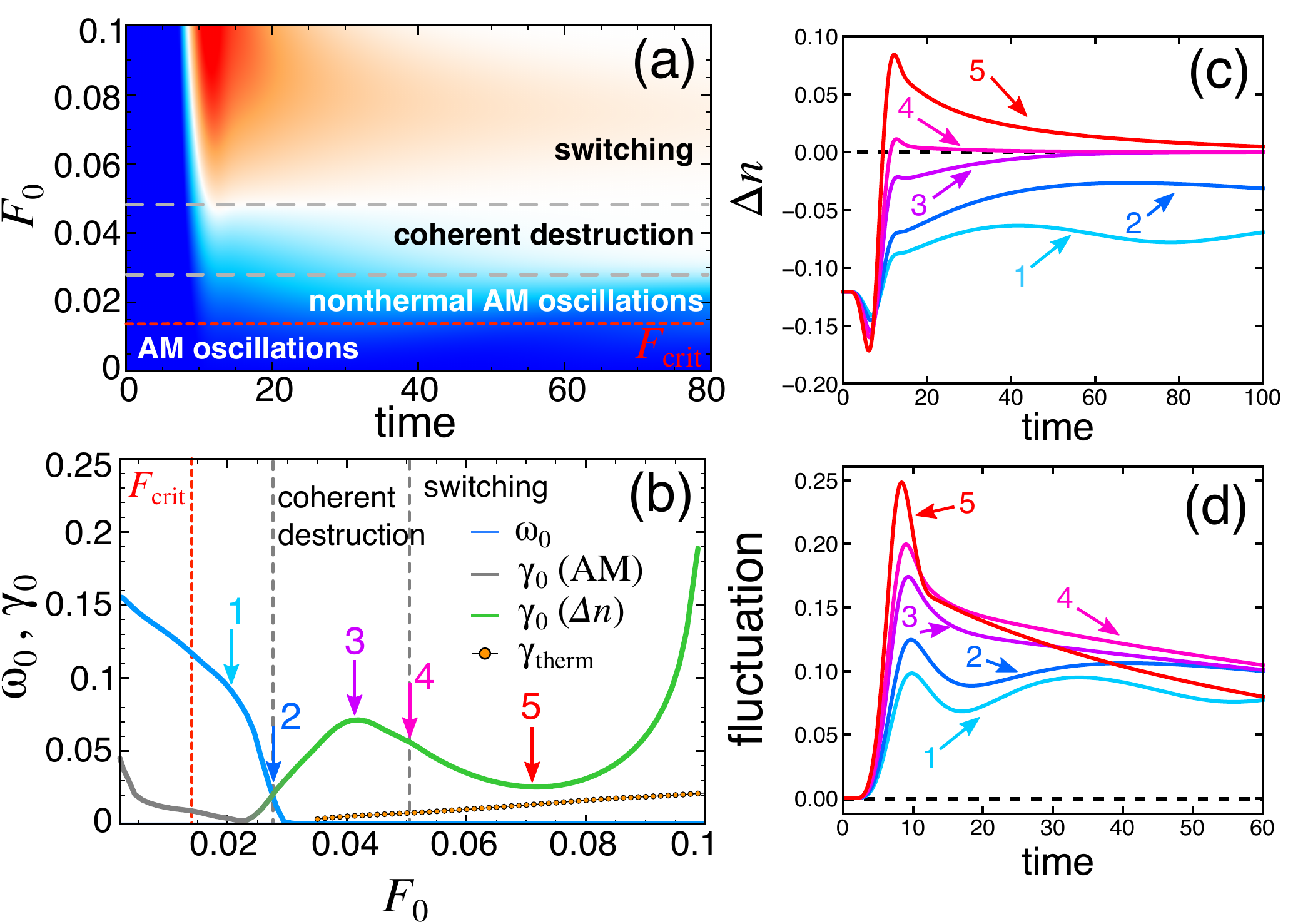}
  \caption{(a) Pulse-induced 2Bloc dynamics of the order parameter for
    varying $F_0$. (b) Oscillation frequency
    $\omega_0$ and 
    the AMs (gray) or the non-oscillatory decay (green) 
    the order parameter. Also shown is the thermalization rate
    $\gamma_\mathrm{therm}$ extracted from $\mathcal{F}(t)$. (c) Order
    parameter for special values of $F_0$ as indicated in (b). (d)
    Fluctuation measure $\mathcal{F}(t)$ at the same values of $F_0$
    as in (c).  \label{fig:2bneq}}
\end{figure}

Performing the analogous scan over the field strength $F_0$ with the
same pulse shape as for the MF case yields the time-dependent order
parameter and the nonequilibrium phase diagram presented in
Fig.~\ref{fig:2bneq}(a) and (b), respectively.  Similarly to the MF
case, the dynamics induced by the SCP exhibits qualitatively distinct
regimes as $F_0$ is increased. For small amplitudes, the order
parameter is subject to AM oscillations.  Increasing the field
strength, the total energy after the pump exceeds the energy of the
equilibrium system at the critical temperature at the value
$F_\mathrm{crit}\simeq 0.014$.  Therefore, for $F> F_\mathrm{crit}$,
one expects complete melting of the CDW after thermalization.
Nevertheless, for $F_\mathrm{crit}<F_0<0.0275$ one finds long-lived AM
oscillations with a small decay rate $\gamma_0$ (see
Fig.~\ref{fig:2bneq}(c)). Such a persistence of nonthermal behavior
over a long time span has also been observed in the dynamics of
antiferromagnetic
order~\cite{werner_nonthermal_2012,tsuji_nonthermal_2013}.

To distinguish the effect of e--e scattering and thermalization from
the effect of coherently precessing pseudospins, it is useful to
introduce the fluctuation measure
$\mathcal{F} \equiv (1/N_k) \sqrt{\sum_{k\sigma}|\langle
  \hat{S}^y_{k\sigma} \rangle|^2}$. This quantity is related to the
magnitude of the $k$-resolved current between the two sublattice sites
and hence provides a convenient measure of a nonthermal state:
$\mathcal{F} = 0$ corresponds to thermal equilibrium, while
$\mathcal{F} > 0$ indicates a nonequilibrium state.
Progressing thermalization can be tracked by the 
decay of $\mathcal{F}(t)$.
The nonthermal AM regime is thus characterized by an extremely
slow thermalization, underpinned by the oscillating
behavior of $\mathcal{F}(t)$ around a nonzero value (Fig.~\ref{fig:2bneq}(d)).

For larger field strength, the order is destroyed rapidly, similarly
as in the MF case. The CD is bounded by two special points: (i) a
point of critical slowing-down at $F_0\simeq 0.0275$ where the
oscillation frequency approaches zero and (ii) the "sweet spot"
$F_0\simeq 0.05$ where the CDW is rapidly destroyed. In contrast to
the MF scenario, no (damped) AM oscillations occur; instead a
non-oscillatory decay of $\Delta n$ is observed. 
The decay rate exhibits a maximum in the middle of the CD regime,
similarly to the MF analysis.
Since e--e collisions facilitate the suppression of the order, the CD
regime is more extended. Note that the thermalization time is still
rather long (Fig.~\ref{fig:2bneq}(d)).

As in the MF analysis, a switching regime appears for larger field
strength, though the order decays due to scattering processes, leading
to additional damping (and almost completely suppressed AM
oscillations). This is due to the larger energy absorption, 
which acts against the CDW order. This is corroborated by the
exponential decay of $\mathcal{F}(t)$ quantified by the thermalization
rate $\gamma_\mathrm{therm}$ (shown in Fig.~\ref{fig:2bneq}(b)).
Nevertheless, there is a pronounced minimum of the decay
rate $\gamma_0$ at $F_0\simeq 0.071$ in the middle of the switching regime.  Here
one encounters a nontrivial situation where, despite the increased
$E_\mathrm{abs}$, the relaxation of the order becomes slower
(dynamical slowing-down).  As the monotonic dependence of
$\gamma_\mathrm{therm}$ shows, this slow-down is a nonthermal
effect. It can be
interpreted as a ``trapping" in a state that is  close to the MF transient state. A
more detailed analysis also shows small oscillations on top of the
exponential decay, which is reminiscent of the AM oscillations in the
switching regime in Fig.~\ref{fig:intro}(c).

\begin{figure}[t]
  \includegraphics[width=\columnwidth]{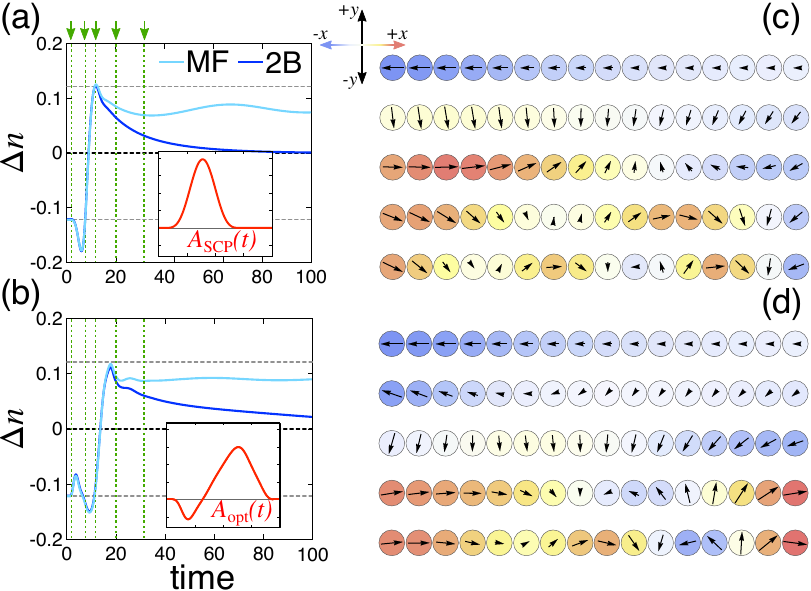}
  \caption{Switching dynamics by a SCP (a) and optimized pulse
    (b). The
    pseudospin configuration for the time points marked with green
    arrows in the small range $[-0.5 \pi/a,-0.4375 \pi/a]$ close the
    Fermi edge is shown for (c) the SCP and (d) the optimized
    pulse. \label{fig:spins}}
\end{figure}

Now, we show how QOCT can stabilize nonthermal states further. In an
optimal switching protocol, the requirements of minimal energy
absorption and switching efficiency need to be balanced.
Using QOCT,
we optimize the fields $A(t)$ to reach a target value of the order
parameter, while simultaneously minimizing $E_\mathrm{abs}$.  Due to
the substantial numerical cost, this procedure is applied only on the
MF level. After identifying optimal pulses~\cite{Note3},
we can then compare the MF and correlated dynamics.  A result of this
optimization procedure is shown in Fig.~\ref{fig:spins}(b).  The
short-time 2Bloc dynamics is well described by the MF approximation.
As compared to the simple SCP (Fig.~\ref{fig:spins}(a)), the switching
efficiency is bigger ($\Delta \bar{n}\simeq 0.75 \Delta n_0$).  In the
MF results, the small amplitude of the AM oscillations indicates the
stability of the state.

It is interesting to analyze the switching dynamics in the pseudospin
picture. While the SCP rotates the spins close to the edge of the
reduced Brillouin zone (Fermi points) by $\pi$ monotonically
(Fig.~\ref{fig:spins}(c)), the optimized pulse initially leads to a
rotation 
in the opposite direction to the switching 
(Fig.~\ref{fig:spins}(d)). As the snapshots of the $k$-dependent
pseudospin configuration demonstrate, this procedure partially
compensates the dephasing and thus results in a better switching.

Including the e--e scattering, $\Delta n(t)$ exhibits a damped
behavior, whereas the short-time dynamics is well caputered by the
MF dynamics.  Albeit $E_\mathrm{abs}$ exceeds the energy difference to
the unordered state, the order parameter decays very slowly. 
Hence, the nonthermal flipped ordered state discussed in Fig.~\ref{fig:2bneq} is -- due
to minimizing $E_\mathrm{abs}$ -- much more pronounced as compared to
switching by the SCP. In fact, $\Delta n(t)$ shows more prominent oscillations on top
of the decay for shorter times.
Especially the pronounced shoulder right after the pulse indicates
that the dynamics is very close to the MF time evolution at short
times. This is a clear signature of the switching regime in
Fig.~\ref{fig:intro}(c). 
Furthermore, the thermalization rate is reduced ($\gamma_\mathrm{therm}=0.0188$ for the SCP,
$\gamma_\mathrm{therm}=0.0132$ for the optimized pulse).

Our pulse optimization can also be applied to the
suppression of the CDW with minimal energy absorption. 
This yields a similarly efficient CD
as for the SCP (cf. Fig.~\ref{fig:2bneq}) but with
reduced $E_\mathrm{abs}$. Even though after complete thermalization
the system ends up in the disordered phase, the fluctuations decay on
the very long
time scale of $\gamma^{-1}_\mathrm{therm} \simeq 170$.

In summary, we revealed the rich \emph{nonequilibrium phase
  diagram} of correlation-driven CDWs after laser excitation,
which includes nonthermal AM oscillations, a sweet spot of rapid destruction of the order, a 
dynamical slow-down of the decay in the switching regime and hence multiple dynamical phase transitions. 
These features are well explained by the MF treatment -- which is rationalized
by the pseudospin picture -- and persist when including
e--e scattering. 
The ability to switch the order or to completely suppress it is a
clear fingerprint of correlation-driven CDWs. Purely phonon-driven CDW
order, on the other hand, exhibits persistent oscillations after
similar laser pulses. Therefore, the transient dynamics and nonthermal
critical behavior allow us to identify the mechanism underlying the
formation of a CDW. 
Furthermore, QOCT allows us to identify
optimal pulses for switching or melting of the order.  Minimizing the
absorbed energy delays the thermalization and may enable the emergence
of competing orders on intermediate timescales.  Controlled access to
long-lived transient states has already resulted in technological
applications~\cite{vaskivskyi_controlling_2015} and we believe that
our approach to QOCT can be usefully applied to the manipulation of a
broad range of electronic orders.

\begin{acknowledgments}
{\it Acknowledgments --} 
  The calculations have been performed on the Beo04 cluster at the
  University of Fribourg and the Piz Daint cluster at the Swiss
  National Supercomputing Centre. This work has been supported by the
  Swiss National Science Foundation through NCCR MARVEL and ERC
  Consolidator Grant No.~724103. We thank Denis Gole\v{z} for fruitful
  discussions.
\end{acknowledgments}

\appendix

\section{Technical details and numerical considerations\label{sec:details}}

\subsection{Hamiltonian and basis representation}

\begin{figure}[h]
  \centering
  \includegraphics[width=0.9\columnwidth]{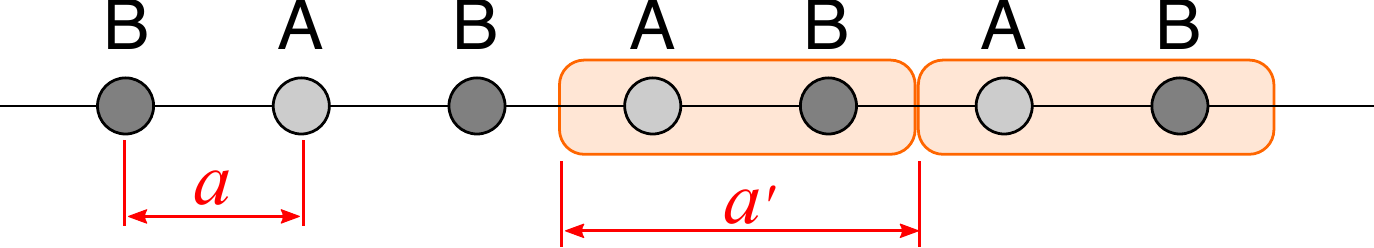}
  \caption{Illustration of the choice of the non-primitive unit cell 
    (lattice constant $a^\prime = 2 a$) in one dimension. \label{fig:unitcell}}
\end{figure}

In order to describe charge-density wave (CDW) states in one dimension, we
introduce an extended unit cell containing two lattice sites, as
sketched in Fig.~\ref{fig:unitcell}. 

In what follows, let us denote the lattice sites of each unit cell by
indices $i,j,\dots$, while the sites in each unit cell will be
labelled by $\alpha,\beta$.  In general, CDW order is driven by two
(possibly coexisting) mechanisms: electron-electron (e--e) interactions
and electron-phonon (e--ph) interactions. In order to take both of them into
account, we extend the Hubbard model
considered in the main text to the Hubbard-Holstein model. Using the above
convention, the Hamiltonian is expressed as
\begin{align}
  \label{eq:ham0}
  \hat{H}(t) = \hat{H}_\mathrm{0}(t) + \hat{H}_{\mathrm{e-e}} +
  \hat{H}_{\mathrm{e-ph}} + \hat{H}_\mathrm{ph} \ ,
\end{align}
where
\begin{align}
  \label{eq:ham1}
  \hat{H}_\mathrm{0} (t) = \sum_{\langle i,j\rangle} \sum_{\alpha \beta}
  \sum_{\sigma }h^{\alpha
  \beta}_{ij}(t) \hdd_{i \alpha \sigma} \hd_{j \beta \sigma} 
\end{align}
denotes the time-dependent electron
Hamiltonian. 
The e--e interaction is modeled by the Hubbard
interaction
\begin{align}
  \hat{H}_\mathrm{e-e} &= \frac{U}{2} \sum_{i,\alpha,\sigma}
  \left(\hdd_{i \alpha \sigma} \hd_{i\alpha \sigma} - \frac12 \right) 
  \left(\hdd_{i \alpha \bar{\sigma}} \hd_{i\alpha \bar{\sigma}} -
  \frac12 \right) \nonumber \\  &\equiv  \frac{U}{2} \sum_{i,\alpha,\sigma}
  \left(\hat{n}_{i \alpha \sigma} - \frac12 \right) 
  \left(\hat{n}_{i \alpha \bar{\sigma}} -
  \frac12 \right)\ .
\end{align}
As usual, $\bar{\sigma} \!= \downarrow$ ($\bar{\sigma} \!= \uparrow$) for
$\sigma \!= \uparrow$ ($\sigma \!= \downarrow$). Furthermore, we account
for Holstein-type phonons described by 
\begin{align}
  \hat{H}_\mathrm{ph} = \omega_\mathrm{ph}\sum_{i,\alpha}
  \hat{b}^\dagger_{i\alpha} \hat{b}_{i\alpha} \ ,
\end{align}
while the electron-phonon interaction is of the form
\begin{align}
  \hat{H}_\mathrm{e-ph} = g \sum_{i,\alpha,\sigma} \hat{n}_{i\alpha
  \sigma} \hat{X}_{i\alpha} \ .
\end{align}
Here,
$\hat{X}_{i\alpha}=[\hat{b}^\dagger_{i\alpha} + \hat{b}_{i\alpha}
]/\sqrt{2}$ represents the phonon distortion. The two limiting cases
of (i) purely correlation-driven and (ii) purely phonon-driven charge
order can be obtained by (i) setting $g=0$, or (ii) $U=0$, respectively.

Transforming to momentum space via
\begin{align}
  \hat{d}_{k\alpha \sigma} = \frac{1}{\sqrt{N}} \sum_m e^{-\iu m k a^\prime}
  \hat{d}_{m\alpha \sigma} \ ,
\end{align}
where $N\rightarrow \infty$ denotes the number of cells, the
Hamiltonian~\eqref{eq:ham1} reads
\begin{align}
  \hat{H}_\mathrm{0} (t) = \sum_{k \in \mathrm{BZ}}\sum_{\alpha \beta} \sum_\sigma
  h_{\alpha \beta}(k;t) \hdd_{k \alpha \sigma} \hd_{k
  \beta \sigma} \ .
\end{align}
Here, the one-body Hamiltonian (in matrix notation) is given by
$\vec{h}(k;t) = \vec{h}^{(0)}(k - A_F(t))$ with
\begin{align}
\vec{h}^{(0)}(k) = 
  \begin{pmatrix} 
    -\mu & -J_0 \left[ 1 + \exp(-\iu k a^\prime) \right] \\ -J_0 \left[ 1 + \exp(\iu k a^\prime) \right] & - \mu 
  \end{pmatrix} \ ,
\end{align}
which has eigenvalues $\en_{\pm}(k) = -\mu \pm 2 J_0 \cos(k a)$. Here,
$A_F(t)$ denotes the electromagnetic vector potential along the chain
direction. Note that the Brillouin zone refers to the extended unit
cell, i.\,e. $\mathrm{BZ}=[-\pi/a^\prime,\pi/a^\prime]$. We assume
half filling ($\mu=0$).

The basis with respect to the sublattice sites (ss) in the extended unit
cell is particularly convenient for introducing approximations to the
e--e interaction, as discussed below. Therefore, all
calculations have been performed in the sublattice representation.
An equivalent basis -- which has been used in the main text -- is
given by the momentum pair $(k,k+Q)$ with the nesting vector
$Q=\pi/a$. The momentum-pair (mp) representation is most useful for defining
pseudospin operators, as introduced in the main text. Any
$k$-dependent matrix in th ss
representation ($\vec{A}_k$) can be transformed to the mp basis by the unitary
transformation
\begin{align}
  \vec{A}^{(\mathrm{mp})}_k = \vec{R}_k
  \vec{A}^{(\mathrm{ss})}_k \vec{R}^\dagger_k 
\end{align}
with
\begin{align}
  \vec{R}_k  = \frac{1}{\sqrt{2}}\begin{pmatrix}
    e^{\iu k a/2} & e^{-\iu k a/2} \\ e^{\iu k a/2} & -e^{-\iu k a/2}
    \end{pmatrix} \ .
\end{align}

\subsection{Mean-field treatment}

Within the mean-field (MF) approximation, the time-dependent
Hamiltonian~\eqref{eq:ham0} is replaced by 
\begin{align}
  \hat{H}^\mathrm{MF}(t) = \sum_{k \in \mathrm{BZ}}\sum_{\alpha \beta} \sum_\sigma
  h^{\mathrm{MF}}_{\alpha \beta}(k-\mathrm{A}_F(t)) \hdd_{k \alpha \sigma} \hd_{k
  \beta \sigma} \ ,
\end{align}
where the one-body MF Hamiltonian reads
\begin{align}
\label{eq:ham_mf}
\vec{h}^{\mathrm{MF}}(k) &= \vec{h}^{(0)}(k) + U
  \begin{pmatrix} 
   \langle \hat{n}_{\mathrm{A}} \rangle -\frac12  & 0\\ 0 & \langle \hat{n}_\mathrm{B} \rangle - \frac12
 \end{pmatrix} 
  \nonumber \\
    &\quad + g
      \begin{pmatrix}
        \langle \hat{X}_\mathrm{A} \rangle & 0 \\
        0 & \langle \hat{X}_\mathrm{B} \rangle 
      \end{pmatrix}
      \ .
\end{align}
Here, $\langle \hat{n}_\mathrm{A} \rangle$
($\langle \hat{n}_\mathrm{B} \rangle$) denotes the occupation on
sublattice site A (B). We assume the paramagnetic case
$\langle \hat{n}_\mathrm{A,B} \rangle = \langle
\hat{n}_{\mathrm{A,B},\uparrow} \rangle = \langle
\hat{n}_{\mathrm{A,B},\downarrow} \rangle $ and therefore drop the spin
indices. $\hat{X}_{\mathrm{A,B}}$ measures the lattice distortion
(corresponding to $k=0$) at the respective sublattice sites. We
introduce the order parameter for CDW
\begin{align}
  \Delta n = \langle\hat{n}_\mathrm{A}\rangle-\langle\hat{n}_\mathrm{B}\rangle
\end{align}
and the distortion parameter
$\Delta X = \langle \hat{X}_\mathrm{A} \rangle -\langle
\hat{X}_\mathrm{B} \rangle$. Expressing the MF
Hamiltonian~\eqref{eq:ham_mf} with these definitions and transforming
to the mp basis then gives the MF Hamiltonian 
\begin{align}
  \hat{H}^{\mathrm{MF}}(t) = \sum_{k \in \mathrm{BZ}} \sum_{\sigma}
  \hat{\vec{c}}^\dagger_{k \sigma} \,\widetilde{\vec{h}}^{\mathrm{MF}}(k,t) \hat{\vec{c}}_{k \sigma}
\end{align}
with $\hat{\vec{c}}_{k \sigma} = (\hc_{k\sigma},\hc_{k+Q,\sigma})$ and 
\begin{align}
  \label{eq:ham_mf_mp}
  \widetilde{\vec{h}}^{\mathrm{MF}}(k,t) = \begin{pmatrix}
    \en_k(t) & \frac{g}{2} \Delta X(t) + \frac{U}{2} \Delta n(t)  \\ 
    \frac{g}{2} \Delta X(t) + \frac{U}{2} \Delta n(t) & \en_{k+Q}(t) 
  \end{pmatrix}\ .
\end{align}
Here, $\en_k(t) = \en_{k-A_F(t)}$ with $\en_k=-2 J_0 \cos(k a)$
denoting the free dispersion. The pseudospin representation can now be
introduced in terms of the Pauli matrices $\gvec{\sigma}^\alpha$ by 
\begin{align}
  \hat{S}^\alpha_{k \sigma} = \frac12 \begin{pmatrix} \hc^\dagger_{k\sigma}
    \hc^\dagger_{k+Q,\sigma} \end{pmatrix}
  \gvec{\sigma}^\alpha \begin{pmatrix} \hc_{k\sigma} \\
    \hc_{k+Q,\sigma} 
  \end{pmatrix}\ .
\end{align}
Exploiting $\en_{k+Q}=-\en_k$, the Hamiltonian
can be written as
\begin{align}
  \hat{H}^{\mathrm{MF}}(t) =\sum_{k\in\mathrm{BZ}} \sum_\sigma \left(
  B^x_{k}(t)\hat{S}^x_{k\sigma} + B^z_{k}(t)\hat{S}^z_{k\sigma} \right)\ ,
\end{align}
where the effective magnetic fields are given by
\begin{align}
  B^x_k(t) = g \Delta X(t) + U \Delta n(t)  \ , \ B^z_k(t) = \en_k(t)
  - \en_{k+Q}(t) \ .
\end{align}

The time-dependent MF equations are solved in two steps. First, the
one-body density matrix in thermal equilibrium $\gvec{\rho}_\mathrm{eq} (k)$ is self-consistently
computed by diagonalizing Eq.~\eqref{eq:ham_mf} and calculating the
order $\Delta n$ and distortion $\Delta X =
-(2g/\omega_\mathrm{ph})\Delta n$ parameters, until
convergence is reached. 
Using $\gvec{\rho}(k,t=0)=\gvec{\rho}_\mathrm{eq} (k)$ as
initial condition the
time evolution is determined by the time stepping 
\begin{align}
  \gvec{\rho}(k,t+\Delta t) = \vec{U}_k(t+\Delta t,t) \gvec{\rho}(k,t)
  \vec{U}^\dagger_k(t+\Delta t,t) \ .
\end{align}
Here, $\vec{U}_k(t+\Delta t,t)$ denotes the time evolution operator of
the MF Hamiltonian, which is computed by fourth-order commutator-free
matrix exponentials~\cite{alvermann_high-order_2011}.

To determine the self-consistent MF Hamiltonian, the phonon
amplitudes $\langle \hat{X}_{\mathrm{A,B}}(t) \rangle$ also need to be
propagated. Combing their respective equations of motion,
the distortion parameter is obtained from the equation
\begin{align}
\label{eq:osceq}
  \frac{1}{2\omega_\mathrm{ph}}\left[\frac{\dd^2}{\dd t^2} +
  \omega^2_\mathrm{ph}\right] \Delta X(t) = - g \Delta n(t) \ ,
\end{align}
which we solve using the fourth-order Numerov
method with the initial condition $\frac{\dd}{\dd t}\Delta X(t) = 0$
for $t=0$.

Since the values $\Delta X(t+\Delta t)$ and $\Delta n(t+\Delta t)$
are needed to carry out the step from $t$ to $t+\Delta t$, the
time propagation is performed in a predictor-corrector fashion.

\subsection{Solution of the Kadanoff-Baym equations \label{subsec:kbe}}

Treating the electron-electron interactions up to second order in $U$
is accomplished by the second-Born (2B) approximation to the
self-energy. In the original lattice representation, the 2B
self-energy reads
\begin{align}
  \label{eq:2b}
  \Sigma_{ij\sigma}(z,z^\prime) = U^2 G_{ij \sigma}(z,z^\prime)
  G_{ij\bar{\sigma}}(z,z^\prime) G_{ji \bar{\sigma}}(z^\prime, z) \ .
\end{align}
Here, $z,z^\prime$ denote the arguments on the Matsubara-Keldysh
contour $\mathcal{C}$, and $G_{ij\sigma}$ stands for the one-body
Green's function
\begin{align}
  G_{ij\sigma}(z,z^\prime) = -\iu \langle \mathcal{T} \hd_{i\sigma}(z)
  \hdd_{j\sigma}(z^\prime) \rangle \ ,
\end{align}
with $\mathcal{T}$ the contour-ordering operator. More details on
the formalism can be found, for instance, in Ref.~\cite{stefanucci_nonequilibrium_2013}.

As discussed in the main text, we employ the local second-Born (2Bloc)
approximation, which is obtained from Eq.~\eqref{eq:2b} by replacing
the index pair $(ij)$ by the diagonal $(ii)$. Switching to $k$-space,
the 2Bloc is cast into a momentum-independent self-energy, 
\begin{align}
  \Sigma_{\alpha \alpha \sigma}(z,z^\prime) = U^2
  \mathcal{G}_{\alpha\alpha\sigma}(z,z^\prime) 
  \mathcal{G}_{\alpha\alpha\bar\sigma}(z,z^\prime)
  \mathcal{G}_{\alpha\alpha\bar\sigma}(z^\prime,z) \ ,
\end{align}
where the index $\alpha$ refers to the sublattice site basis and
\begin{align}
  \mathcal{G}_{\alpha\beta\sigma}(z,z^\prime) =
  \frac{1}{|\mathrm{BZ}|} \int_{\mathrm{BZ}} \dd k \, G_{\alpha \beta\sigma}(k; z,z^\prime)
\end{align}
defines the local Green's function.

Switching to a matrix notation for the sublattice indices (and dropping
the spin index), the equation
of motion for the Green's function assumes the standard form
\begin{align}
  \left[\iu \partial_z - \vec{h}^\mathrm{MF}(z) \right]
  \vec{G}(k;z,z^\prime) = \int_\mathcal{C}\!\dd z^\dprime \,
  \gvec{\Sigma}(z,z^\dprime) \vec{G}(k; z^\dprime,z^\prime) \ .
\end{align}
Projecting onto imaginary and real times and invoking the Langreth
rules then yields the Kadanoff-Baym equations (KBEs). The KBEs are
solved using an in-house massively parallel computer code based on a
fourth-order implicit predictor-corrector algorithm. For the results
presented in the main text, the time interval was split into
$N_t=3000$ equidistant points, while the imaginary branch (for the
nonequilibrium calculations) was represented by $N_\tau=800$ grid
points. The Green's function for every $k$-point has to be propagated
simultaneously, which is accomplished by a distributed-memory
layout. For obtaining converged results, we used $N_k=256$ points in
the Brillouin zone.

\subsection{Equilibrium spectral function and mean-field fit}

\begin{figure}[t]
  \includegraphics[width=\columnwidth]{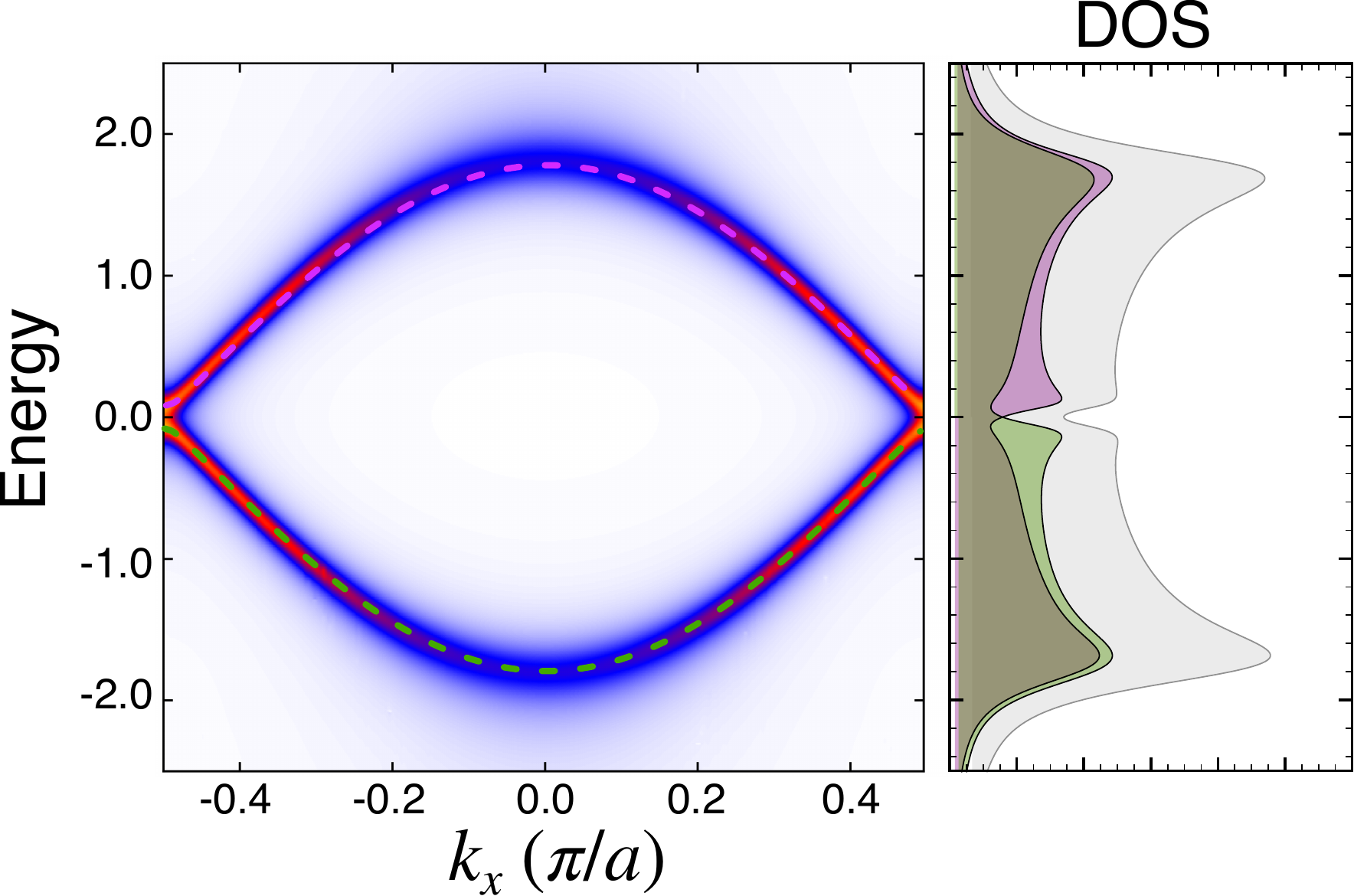}
  \caption{Left panel: spectral function $A(k,\omega)$ (summed over bands) within
    the 2Bloc approximation. The dashed lines represent the fit by
    the MF model with renormalized parameters $\widetilde{J}_0$ and
    $\widetilde{U}$. Right panel: band-resolved 
    (green and purple filled curves) and total (gray filled curve)
    density of states within the 2Bloc approximation.  The parameters are, as in the
    main text, $U=-2$, $\beta=40$.
    \label{fig:spectral}}
\end{figure}

Before the KBEs can be solved (see subsec.~\ref{subsec:kbe}), the
equilibrium (Matsubara) Green's function needs to be
computed. To this end, we solve the corresponding Dyson equation
\begin{align}
\label{eq:dyson}
  \vec{G}(k;\tau) &= \vec{g}(k;\tau) \nonumber \\ &\quad + \int^\beta_0\! \dd \tau^\prime
  \int^\beta_0\! \dd \tau^\dprime \, \vec{g}(k;\tau-\tau^\prime)
  \gvec{\Sigma}(\tau^\prime-\tau^\dprime) \vec{G}(k;\tau^\dprime) \ .
\end{align}
Here, $\vec{g}(k,\tau)$ denotes the MF Green's function, while
$\gvec{\Sigma}(\tau)$ is the self-energy in the 2Bloc approximation. The
Dyson equation~\eqref{eq:dyson} is solved by a combination of Fourier
transformation and fifth-order fix-point iteration to improve the
accuracy. A description of the method can be found in
Ref.~\cite{schuler_spectral_2017}. As for the nonequilibrium
calculations, we use $N_k=256$ $k$-points, whereas $N_\tau=4096$
points on the Matsubara axis were needed for converging the
results. For the nonequilibrium calculations, the Matsubara Green's
function is defined on the reduced imaginary grid by interpolation. 

The spectral function $\vec{A}(k,\omega)$ in real-frequency space is
obtained by Pad\'e analytic continuation as in
Ref.~\cite{schuler_spectral_2017}. The band-integrated spectral
function $A(k,\omega)=\sum_\alpha A_{\alpha \alpha}(k;\omega)$ is
shown in Fig.~\ref{fig:spectral}. In accordance with the Luttinger-Ward
theorem, the broadening due to many-body effects is least pronounced
in the vicinity of the chemical potential $\mu =0$, while significant
broadening is apparent at the band top and bottom. Since we are in the 
weak-coupling regime, the main effect of the electronic
correlations is a renormalization of the bands. 

In order to be able to directly compare the dynamics within the MF and
2Bloc approximation, the band renormalization is taken into account in
the effective parameters of the MF Hamiltonian~\eqref{eq:ham_mf},
replacing $J_0 \rightarrow \widetilde{J}_0$ and $U \rightarrow
\widetilde{U}$. These parameters are determined by fitting the MF band
structure to the maximum (with respect to $\omega$) of $A(k,\omega)$,
while requiring the order parameter to be identical (see Fig.~\ref{fig:spectral}). The result is
$\widetilde{J}_0=0.89 J_0$ and $\widetilde{U}=0.625 U$. The good
quality of the fit is supported by the almost identical short-time
dynamics within the MF and 2Bloc approximation, ensuring that applying
the QOCT on the MF level provides optimal pulses for the correlated
dynamics, as well.

\subsection{Quantum optimal control\label{subsec:qoct}}

The optimization of the laser pulses by quantum optimal control theory
(QOCT) is performed on the MF level. Since the MF dynamics is
described by a nonlinear equation of motion for the one-body density
matrix, the usual approach based on an (effective) Schr\"odinger
equation (Krotov algorithm) is not applicable. In fact, one has to
resort to gradient-free optimization methods because the derivative
with respect to the driving field can not be obtained analytically. 

One can expect that pulses containing a minimal amount of mean field
energy $E_\mathrm{p}=(\epsilon_0/T_\mathrm{p}) \int \dd t \,
|E_F(t)|^2 = (\epsilon_0/T_\mathrm{p}) \int \dd t \, |\dot{A}_F(t)|^2$
-- as
required to minimize
heating effects 
-- are relatively smooth functions without strong variations. 
On the other hand, the search space
has to be large enough to find good approximations to the optimal
fields. To fulfil these objectives, we parameterize the vector
potential by
\begin{align}
  A_F(t) = \sum^{N_b}_{i=1} c_i B_i(t) \ ,
\end{align}
where $B_i(t)$ are fourth-order B-splines with respect to the time
interval $[t_0,t_0+T_\mathrm{p}]$. 
To ensure the corresponding
electric field $E_F(t) = - \dot{A}_F(t)$ is zero at the end points of
the interval and make sure no momentum is transferred to the system
($A_F(t_0+T_\mathrm{p})=A_F(t_0)=0$), the boundary coefficients are
fixed by $c_1=c_2=c_{N_b-1}=c_{N_b}=0$. 

For the switching scenario, we are interested in long-time stable
dynamics of $\Delta n(t)$. As is known from the analysis with respect to
the single-cycle pulses, amplitude mode oscillations are expected to
be present around a switched value of the order parameter after time $t_1$. 
We thus perform 
a linear fit $\Delta n_\mathrm{fit}(t) = a (t-t_1) +b$ to the
dynamics of $\Delta n(t)$ after the pulse. 
We then required
that (i) the mean value of the order, encoded in $b$, is maximal,
while (ii) the average slope $|a|$ should be minimal. The condition
(ii) is necessary for the long-time stability of the switched state to
ensure no drift can occur at longer time scales. Similarly, for coherent
destruction one requires $|b|$ to be minimal. Gathering the B-spline
coefficients in the vector $\vec{c}$, the target functional for
switching the order from $\Delta n(t=0)<0$ is given by
\begin{align}
  \label{eq:func1}
  J_\mathrm{switch}[\vec{c}] = -b + \epsilon_1 |a| + \epsilon_2 E_\mathrm{abs} \ ,
\end{align}
while we use
\begin{align}
  \label{eq:func2}
  J_\mathrm{CD}[\vec{c}] = |b| + \epsilon_1 |a| + \epsilon_2 E_\mathrm{abs}
\end{align}
for achieving coherent destruction. Here, $\epsilon_1$ is a penalty parameter for the slope, whereas
$\epsilon_2$ denotes the penalty with respect to the absorbed energy
$E_\mathrm{abs}$. In order to evaluate the functionals~\eqref{eq:func1}
or \eqref{eq:func2}, one has to perform the time propagation up to a
sufficiently large time $T_\mathrm{max}$ (we set
$T_\mathrm{max}=500$), compute the fitting parameters $a$, $b$ and the
absorbed energy. As mentioned above, the gradient with respect to
$\vec{c}$ cannot be calculated directly. Therefore, we minimize the
functionals by a combination of the Pikaia genetic
algorithm~\cite{charbonneau_genetic_1995} for finding local minima and
the NEWUOA algoritm~\cite{pillo_large-scale_2006} for finding the
global minimum. This procedure depends on the parameters $N_b$,
$\epsilon_1$, $\epsilon_2$ and the pulse duration $T_\mathrm{p}$.

\section{Dynamics of phonon-driven charge order}

\begin{figure}[t]
  \includegraphics[width=0.9\columnwidth]{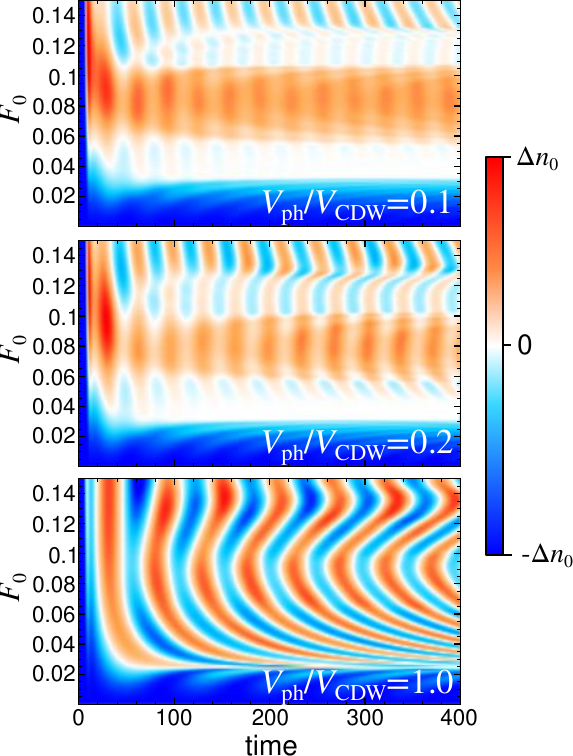}
  \caption{Dynamics of the order parameter $\Delta n(t)$ induced by
    single-cycle pulses with strength $F_0$ (analogous to Fig.~1(b) in
    the main text), for different ratios of the contribution to the CDW
    from  e--ph interactions (we fix $\omega_\mathrm{ph}=0.2$).\label{fig:phonon}  }
\end{figure}

\begin{figure*}[t]
  \includegraphics[width=0.9\textwidth]{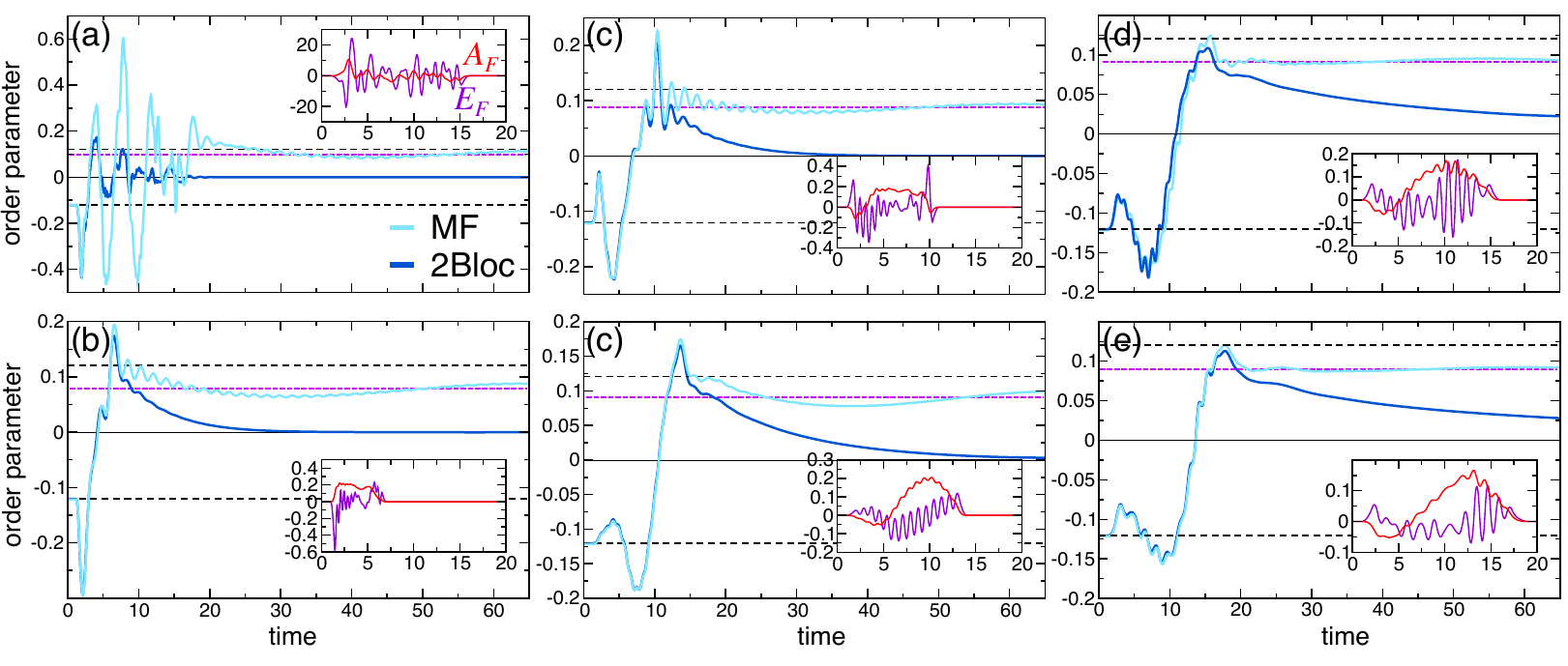}
  \caption{Switching dynamics of the CDW order parameter induced by
    selected optimized pulses in the MF (light blue) and 2Bloc (dark blue)
    approximation. The panels (a)--(e) correspond to increasing
    energy penalty $\epsilon_2$; in (a) $\epsilon_2=0$. The black
    dashed lines indicate the initial and flipped value $\pm\Delta
    n_0$, while the purple short-dashed lines represents the mean
    value $\Delta \bar{n}$ within the MF approximation.\label{fig:switching}  }
\end{figure*}

As discussed in the main text, the distinct regimes in the
nonequilibrium phase diagram are tightly connected to the driving
mechanism of the CDW.  In order to corroborate this behavior and,
moreover, test the robustness of nonequilibrium features with respect
to lattice distortions, we included e-ph couplings. Both,
the e--e and the e--ph coupling is responsible
for the formation of the CDW. The relative contribution from either
effect can be captured by the parameter
\begin{align}
  \label{eq:vcdw}
  V_\mathrm{CDW} \equiv V_\mathrm{ph} + V_\mathrm{e}  =
  \frac{g^2}{\omega_\mathrm{ph}} - \frac{U}{2} \ .
\end{align}
The parameter~\eqref{eq:vcdw} is related to the interaction energy, as
can be seen by computing the total energy in the MF approximation:
\begin{align*}
  E_\mathrm{tot} &= 2\sum_{k} \mathrm{Tr} \left[ \vec{h}^{(0)}(k)
  \gvec{\rho}(k)\right] + \sum_{k} \mathrm{Tr} \left[ (\vec{h}^{\mathrm{MF}}(k)-\vec{h}^{(0)}(k))
  \gvec{\rho}(k)\right] \\ &\equiv E_0 + E_\mathrm{int} \ .
\end{align*}
Expressing the interaction energy by the order and distortion
parameters in equilibrium, one finds
$E_\mathrm{int} = V_\mathrm{CDW} \Delta n ^2$.  Note that an identical
value of $V_\mathrm{CDW}$, regardless of the individual
contributations of the e--e or e--ph interactions, gives rise to the
same value of $\Delta n$ and the gap size.

Fixing $V_\mathrm{CDW}=0.625$ (corresponding to the results of the
main text) we now vary the ratio $V_\mathrm{ph}/V_\mathrm{CDW}$ and
study how the increased contribution of e--ph interactions
to the order affects the pulse-induced dynamics.
Figure~\ref{fig:phonon} shows the nonequilibrium phase diagram (MF
approximation) analogous to Fig.~1(b) in the main text for
$V_\mathrm{ph}/V_\mathrm{CDW} = 0.1$ (top),
$V_\mathrm{ph}/V_\mathrm{CDW}=0.2$ (middle) and
$V_\mathrm{ph}/V_\mathrm{CDW}=1$ (bottom). For a CDW dominated by 
e--e
correlation
effects, the different regimes of amplitude mode oscillations,
coherent destruction and switching are qualitatively still present,
but superimposed with coherent phonon oscillations. 
It is interesting
to see that the lower boundary of the coherent destruction regime
represents the fastest way to destroy the order, whereas the ``sweet
spot" is exhibiting more oscillations. In general, the amplitude of the
phonon oscillations increases under stronger driving.

The qualitative behavior of the laser-driven nonequilibrium regimes is
still present for $V_\mathrm{ph}/V_\mathrm{CDW}=0.2$, albeit the
boundaries are smeared out by the phonon oscillations. For an even larger
contribution of the electron-phonon coupling, the dynamics is
dominated by the phonons and thus displays the generic behavior of the
purely phonon-driven case (bottom panel in Fig.~\ref{fig:phonon}). In
this scenario, the persistent oscillations of $\Delta n$ with
frequency $\omega_\mathrm{ph}$ are the
dominating feature. Neither destruction nor switching of the CDW is
possible anymore. 

We conclude that a qualitatively different dynamics
of the order parameter for different pulse amplitudes is a clear signature of a predominantly
correlation-driven CDW formation. Small e--ph coupling leads
to small additional coherent phonon oscillations, but does not suppress the 
characteristic features discussed in the main text. A larger contribution of
the phonons, on the other hand, suppresses any switching behavior.

\section{Switching and coherent destruction by optimized pulses}

As explained in subsec.~\ref{subsec:qoct}, the pulse optimization with
the aim of switching the CDW is performed on the MF level and depends
on the number of B-spline coefficients $N_b$, the slope penalty
$\epsilon_1$, the penalty with respect to the absorbed energy
$\epsilon_2$ and the pulse duration $T_\mathrm{p}$. We performed the pulse
optimization for various combinations of these parameters and found
that $N_b=28$ is enough to find the optimal pulse shapes. Increasing
$N_b$ yields essentially the same pulses with extra oscillations on
top. Furthermore, the value of $\epsilon_1$ affects the pulses only
weakly since most of the resulting pulses result in a vanishing
average slope of $\Delta n(t)$. The pulse length $T_\mathrm{p}$ was
varied from $T_\mathrm{p} = 5.0$ to $T_\mathrm{p} =20.0$; we select
the best pulses in this range for a fixed value of $\epsilon_2$.

\subsection{Switching dynamics}

The energy penalty $\epsilon_2$ is the most crucial
parameter. Choosing $\epsilon_2=0$ results in very strong pulses,
leading to almost perfect switching on the MF level 
(Fig.~\ref{fig:switching}(a)). However, within the 2Bloc approximation, 
the huge amount of absorbed energy rapidly destroys the order. 
Further analysis
shows that the system thermalizes at a very high effective
temperature shortly after the pulse.  Gradually increasing
$\epsilon_2$ decreases the switching efficiency
(Fig.~\ref{fig:switching}(b--e)) while reducing the energy
absorption. This leads to a longer life time of the switched
state within the 2Bloc dynamics. Interestingly, the shape of the
vector potential $A_F(t)$ looks quite similar in
Fig.~\ref{fig:switching}(c)--(e). It corresponds to the minimization
of dephasing which is explained in the main text. The best compromise
between energy absorption and switching is provided by the pulse in
Fig.~\ref{fig:switching}(e). We found that applying a smoothening
low-pass filter further reduces $E_\mathrm{abs}$, while the short-time
dynamics is not altered. This optimal pulse is the one presented and
discussed in the main text. Note that increasing $\epsilon_2$ further
leads to a suppression of switching, since the requirement to minimize
the absorbed energy -- which is zero if no pulse is applied -- starts
to dominate.

\subsection{Coherent destruction dynamics}

\begin{figure}[b]
  \includegraphics[width=\columnwidth]{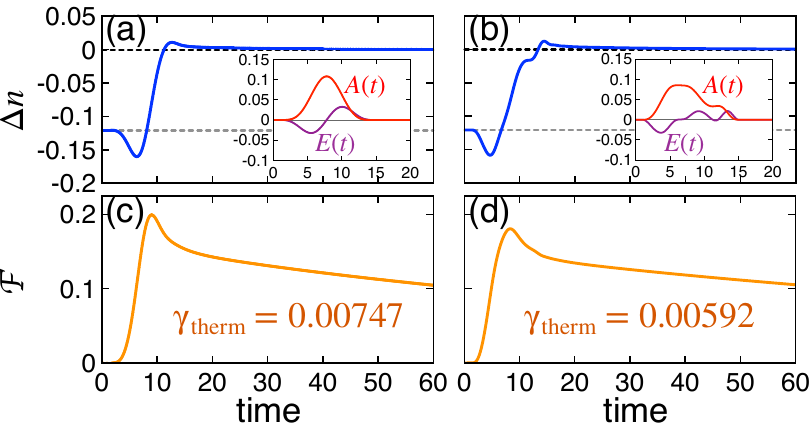}
  \caption{Dynamics on the 2Bloc level at the sweet spot of coherent
    destruction: (a) single-cycle pulse, (b) optimized pulse. The
    insets show the laser fields. The corresponding
    fluctuation measure $\mathcal{F}(t)$ is shown in panels (c) and (d), respectively.\label{fig:cd}  }
\end{figure}

An analogous analysis was carried out for the coherent destruction of
the CDW order. However, we found that the optimal pulse and the
resulting dynamics is very robust against changes of $\epsilon_1$ and
$\epsilon_2$. The pulse with the smallest $E_\mathrm{abs}$ is shown in
Fig.~\ref{fig:cd}(b) and compared to the dynamics driven by the
single-cycle pulse at the "sweet spot" (Fig.~\ref{fig:cd}(a))
discussed in the main text. It is interesting to note that the simple
single-cycle pulse results in perfect suppression of the order while
injecting only little energy into the system. Correspondingly, the
optimized field $A_F(t)$ is qualitatively almost the same as the
single-cycle pulse. However, the absorbed energy is reduced, such that
the thermalization (Fig.~\ref{fig:cd}(d)) is slower than for the
single-cycle pulse (Fig.~\ref{fig:cd}(c)).

\section{Anisotropic two-dimensional lattice}

\begin{figure}[t]
  \includegraphics[width=\columnwidth]{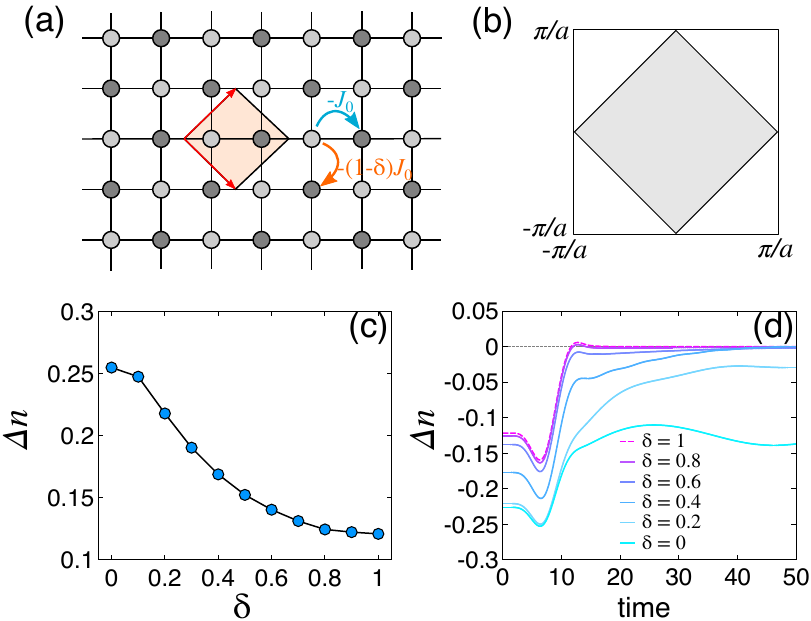}
  \caption{(a) Sketch of the checker-board CDW order in an anisotropic
    2D square lattice. The shaded area shows the unit cell chosen for
    the calculations. (b) Full (while square) and reduced (gray
    shaded) Brillouin zone. (c) Dependence of the order parameter
    $\Delta n$ on the anisotropy $\delta$. (d) Dynamics induced by the
    single-cycle pulse at the point of coherent destruction for
    different values of $\delta$. \label{fig:aniso} }
\end{figure}

In the main text, we consider a one-dimensional configuration of the lattice.  
Note that the local approximation to the self-energy leads to
generic features of a higher-dimensional system, while the 1D
character primarily enters via the free band structure. Most CDW
orders observed in materials are, in fact, two-dimensional (typically
stripe or checker board order).  In this section we confirm that our
results based on the 1D system are also valid for the 2D case with
anisotropic hopping. To be concrete, we consider a square lattice with
hopping $J_0$ in the $x$-direction and $(1-\delta) J_0$ in the
$y$-direction (see Fig.~\ref{fig:aniso}(a)); $\delta=0$ corresponds to
the isotropic 2D system, while $\delta =1$ recovers the 1D limit. The
CDW forming in this configuration follows a checker-board order,
corresponding to a nesting vector $\vec{Q}=(\pi/a,\pi/a)$.

The derivations in Section~\ref{sec:details} are applicable to
the 2D case, as well, after (i) replacing the 1D wave vector $k$ by
a vector $\vec{k}$ from the reduced Brillouin zone shown in
Fig.~\ref{fig:aniso}(b), and (ii) modifying the free Hamiltonian (mp basis) to
\begin{align}
  \vec{h}^{(0)}(\vec{k}) = \begin{pmatrix}
    \varepsilon(\vec{k}) &
    \frac{U}{2}\Delta n \\  \frac{U}{2}\Delta n & \varepsilon(\vec{k}+\vec{Q}) 
  \end{pmatrix} \ .
\end{align}
Here, 
\begin{align}
  \varepsilon(\vec{k}) = -2 J_0 \left(\cos(k_x a) + (1-\delta)\cos(k_y
  a) \right)
\end{align}
denotes the original free band structure.

We have performed equilibrium calculations with the 2Bloc
approximation for different values of $\delta$ (see
Fig.~\ref{fig:aniso}(c)). The order parameter $\Delta n$ deviates by
less than 10\% from the 1D value in the regime of anisotropy $\delta =
0.7\dots 1$. This relatively large span shows that small deviations
from our 1D setup have almost no influence on the results discussed in
the main text. 

Furthermore, we have analyzed the pulse-induced dynamics in the 2D
scenario. As an example, we show the dynamics of the order parameter
at the ``sweet spot" of coherent destruction within the MF approximation in
Fig.~\ref{fig:aniso}(d). We applied the same pulse as for the 1D case
(polarization along the $x$ direction). One observes similar behavior as for the
equilibrium properties: for moderately strong anisotropy, the time
evolution of $\Delta n$ is very close to the 1D case. Therefore, the
different regimes of the nonequilibrium phase diagram discussed in the main text
are also relevant for the anisotropic 2D system.

\end{document}